\documentclass{iau}
\usepackage{graphicx}

\title[Monte Carlo modelling of globular star clusters] 
{Monte Carlo modelling of globular star clusters \textendash \  many primordial binaries, IMBH formation}

\author[Mirek Giersz, Nathan Leigh, Michael Marks, Arkadiusz Hypki \& Abbas Askar]   
{Mirek Giersz$^1$, Nathan Leigh$^{2,3}$, Michael Marks$^4$, Arkadiusz Hypki$^{1,5}$
 \and Abbas Askar$^1$}

\affiliation{$^1$Nicolaus Copernicus Astronomical Centre, Polish Academy of Sciences, \\ ul. Bartycka 18, 00-716 Warsaw, Poland \\ emails: {\tt mig@camk.edu.pl \\ \tt ahypki@camk.edu.pl \\ \tt askar@camk.edu.pl} 
\\[\affilskip]
$^2$Department of Physics, University of Alberta, \\ CCIS 4-183, Edmonton, AB T6G 2E1, Canada \\email: {\tt nleigh@ualberta.ca}
\\[\affilskip]
$^3$Department of Astrophysics, American Museum of Natural History, \\ Central Park West and 79th Street, New York, NY 10024 
\\[\affilskip]
$^4$Helmholtz-Institut f¨ur Strahlen- und Kernphysik, \\ Nussallee 14-16, D-53115, Bonn, Germany \\email: {\tt mmarks@astro.uni-bonn.de}
\\[\affilskip]
$^5$Leiden Observatory, Leiden University, \\ P.O. Box 9513, 2300 RA Leiden, The Netherlands }
\pubyear{2014}
\volume{312}  
\pagerange{119--126}
\jname{Star Clusters and Black Holes in Galaxies Across Cosmic Time}
\editors{A.C. Editor, B.D. Editor \& C.E. Editor, eds.}
\begin{document}

\maketitle

\begin{abstract}
We will discuss the evolution of star clusters with an large initial binary fraction, up to 95\%. The initial binary population is chosen to follow the invariant orbital-parameter distributions suggested by Kroupa (1995). The Monte Carlo MOCCA simulations of star cluster evolution are compared to the observations of Milone et al. (2012) for photometric binaries. It is demonstrated that the observed dependence on cluster mass of both the binary fraction and the ratio of the binary fractions inside and outside of the half mass radius are well recovered by the MOCCA simulations. This is due to a rapid decrease in the initial binary fraction due to the strong density-dependent destruction of wide binaries described by Marks, Kroupa \& Oh (2011). We also discuss a new scenario for the formation of intermediate mass black holes in dense star clusters. In this scenario, intermediate mass black holes are formed as a result of dynamical interactions of hard binaries containing a stellar mass black hole, with other stars and binaries. We will discuss the necessary conditions to initiate the process of intermediate mass black hole formation and the dependence of its mass accretion rate on the global cluster properties.
\end{abstract}

\firstsection 
\section{Introduction}

Recent high resolution observations of globular clusters (GC) provide a very detailed picture of their physical status and show complex phenomena connected with multiple stellar populations, binary evolution and the Galactic tidal field. Despite such great observational progress there are many theoretical uncertainties connected with the origins of GCs and the properties of their primordial binary populations. To bridge the gap between present-day observed binary properties and their properties at the time of cluster formation, we need to discriminate between different theories and models by means of numerical dynamical simulations of GCs. Based on then available observations of the late-type stellar binary population in the Galactic field, Kroupa (1995) suggested that, taking into account dynamical processing, the initial binary population in star clusters is largely invariant, in the sense that almost every star forms in a binary system with invariant formal distribution functions (due to energy and angular momentum conservation, the physics of molecular clouds, all of which are the same everywhere, except perhaps in very intense star bursts). These distribution functions are parent distribution functions, from which a particular case is discretized, or rendered, and their suggested invariance is tightly connected to the notion of an invariant IMF (Kroupa \& Petr-Gotzens 2011). These parent functions can have different properties in different mass ranges (brown dwarfs, late and early type stars). The Kroupa (1995) set-up of the primordial binary populations has been shown to work well for the late-type stellar population (solar-type stars and below) in young star forming regions (Marks \& Kroupa 2012, Marks et al. 2014) and the Galactic field (Marks \& Kroupa 2011). Such distribution functions are needed to initialize N-body models in order to study how young and old clusters evolve into the field and associations. The simulations of GCs described in this talk use the Kroupa (1995) initial binary population and were conducted by the Monte Carlo code MOCCA (Hypki \& Giersz 2013, Giersz et al. 2013 and references therein). The results of the simulations were compared to observational data using the photometric binaries provided in Milone et al. (2012), and to the initial dissolution rate of primordial binaries found using N-body simulations in Marks, Kroupa \& Oh (2011).

The presence of intermediate mass black holes (IMBH) in the cores of some GCs has been debated for a long time. There are many theoretical arguments in favor of the formation of IMBHs in the centers of GCs (e.g. L\"{u}tzgendorf et al. 2013 and references therein), but there is yet no observational confirmation of the presence of any IMBH in any Galactic GCs. All proposed scenarios for the formation of IMBHs in GCs require special initial conditions: 1) the formation of very massive Population III stars (Madau \& Rees 2001), 2) runaway merging of main sequence stars in young and very dense star clusters (Portegies Zwart et al. 2004, G\"{u}rkan et al. 2004), 3) accretion of residual gas on stellar mass black holes (BH) formed from the first generation stars (Leigh et al. 2013a) and 4) tidally stripped parent galaxy cores of neighboring dwarf galaxies (e.g. Baumgardt et al. 2003). A new scenario for IMBH formation is proposed in this talk and does not need very specific initial conditions. An IMBH is built-up only via binary dynamical interactions and mass transfer (also induced by dynamical interactions) in binaries. 

\section{Method}

The MOCCA code used for the star cluster simulations presented here is the Monte Carlo code based on H\'{e}non's implementation of the Monte Carlo method (H\'{e}non 1971), which was further substantially developed by Stod\'{o}\l{}kiewicz in the early eighties (Stodol-\\kiewicz 1986). This method can be regarded as a statistical way of solving the Fokker-Planck equation. A star cluster is treated as a set of spherical shells, each of which represents an individual object: star, binary or a group of the same objects. Each shell is characterized by mass, energy and angular momentum. Relaxation of a given object with all other objects in the system is approximated via the interaction of two neighboring shells. There are two independently developed Monte Carlo codes: by Fred Rasio's group (Morscher et al. 2014 and reference therein) and by my group (Giersz et al. 2013 and reference therein). Actually, there are two more Monte Carlo codes, which were recently developed by Vasiliev (2014) for non spherical stellar systems and by Sollima \& Mastrobuono Battisti (2014) for a realistic treatment of the tidal field. 

The basic assumptions behind the Monte Carlo method are the following: 1) spherical symmetry, which makes it easy to quickly compute the gravitational potential and stellar orbits at any place in the system. That is a very severe assumption, e.g. the evolution of a system with rotation cannot be investigated. In specified spherically symmetric potentials every star is characterized by its mass, energy and angular momentum. Each star moves on a rosette orbit. There is no need for integration of an orbit in the Monte Carlo method, since it is easy to calculate the position of each star along its orbit. In this way, the Monte Carlo method is very fast; 2) dynamical equilibrium, we cannot follow the very initial stages of cluster evolution connected e.g. with primordial gas removal and violent relaxation; 3) cluster evolution is driven by two-body relaxation, and the time step at each position in the system is proportional to the local relaxation time. That is the second reason why the Monte Carlo method is so fast. Generally, we cannot follow any physical processes with characteristic time scales much shorter then the relaxation time.

The MOCCA code (Giersz et al. 2013) is a "kitchen sink" code, which is able to follow most physical processes that are important during star cluster dynamical evolution. For stellar and binary evolution, Jarrod Hurley's BSE code is used (Hurley et al. 2000, Hurley et al. 2002), for the scattering experiments John Fregeau's Fewbody code is used (Fregeau et al. 2004), and the realistic description of escape processes in tidally limited clusters is done on the basis of the Fukushige \& Heggie (2000) theory. The MOCCA code provides as many details as N-body codes. It can follow the time evolution and movement of particular objects. The MOCCA code is extremely fast. It needs about a day to complete the evolution of a realistic globular cluster. So, instead of just one N-body model, hundreds or thousands of models can be computed with different initial conditions. The MOCCA code is ideal either for dynamical models of a particular cluster or for large surveys. 

\section{Results}

The results presented in this talk came as a byproduct from two projects carried out with Nathan Leigh and other collaborators. The aim of the first project (Leigh et al. 2013b) was to explain an observed anti-correlation between cluster mass and binary fraction (Milone et al. 2012) and between the strength of low-mass star depletion in the present-day mass function (MF) and the cluster concentration (De Marchi et al. 2007). The aim of the second project (Leigh et al. 2014) was to constrain the initial properties of primordial binaries by comparison with observations (Milone et al. 2012) and to check if star cluster simulations with initial conditions drawn from the invariant Kroupa (1995) distributions are able to recover the observed spatial distributions of binaries. All together, for these projects about 400 models of star clusters (SC) were simulated by the MOCCA code. 

The model parameters run for the mentioned projects are as follows. Generally, more massive models also have larger concentrations (measured as the ratio between the tidal and half-mass radii -  $R_{plum}=R_t/R_h$), with some of them being extremely concentrated ($R_{plum}$ as large as 125). Most of the initial models had binary fractions equal to 0.1, but for some of them it varied between 0.3 to 0.95. The binary period distribution was uniform in the logarithm of the semi-major axis up to 100AU for all standard models, and up to 200AU and 400AU for all other models. For a substantial number of models, instead of the flat semi-major axis distribution, the Kroupa (1995) period distribution was used, up to log(P)=8.3. Also, different IMFs were used: the Kroupa canonical - a two segmented IMF (Kroupa 2001), the Kroupa standard - a three segmented IMF (Kroupa, Tout \& Gilmore 1993), and the two segmented modified Kroupa IMF (different power-law indexes). Supernovae (SN) natal kick velocities for neutron stars (NS) and BHs were modified for some models according to the mass fallback procedure described by Belczynski et al. (2002). 

\subsection{Extremely Large Initial Binary Fraction}

\subsubsection{Time Evolution of Binary Parameters in MOCCA and N-body Computations}

The results of the MOCCA computations for GCs with the Kroupa (1995) primordial binary population and their comparison with observational data (Milone et al. 2012) are described in detail in Leigh et al. (2014). Here we concentrate on the evolution of the binary parameters in evolving clusters described by tidally filling models (with $R_{plum}$ controlled by the $W_0=6$ King model), and strongly concentrated tidally under-filling $W_0=6$ King models with $R_{plum}=50$. Also, we present a comparison between the MOCCA results and the results of the BiPoS code developed by Michael Marks, based on Marks, Kroupa \& Oh (2011) and Marks \& Kroupa (2011). The BiPoS code offers an analytic description for the processing of the Kroupa (1995) initial binary population seen in N-body computations with initial masses up to $10^{3.5} M_{\odot}$. BiPoS thus evolves the initial binary population efficiently for cluster ages up to 5 Myr, the time for which their computations were run.

\begin{figure}[h]
\vspace*{-0.1 cm}
\begin{center}
 \includegraphics[width=7.0cm, angle=-90]{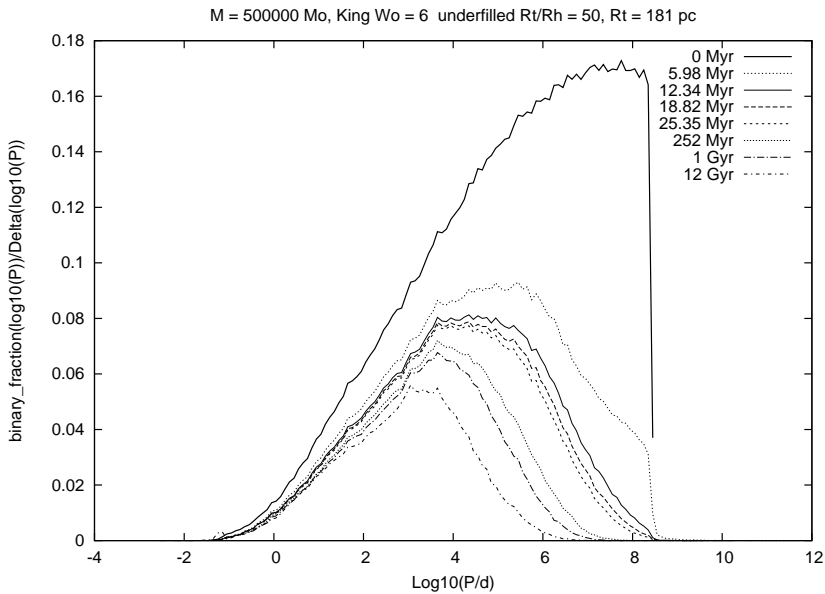} 
 \vspace*{0.8 cm}
 \caption{Time evolution of the binary period distribution in one of our tidally under-filling star cluster models.}
   \label{fig1}
\end{center}
\end{figure} 

Fig. 1 shows the evolution of the primordial binary period distributions for one of our tidally under-filling models. As expected, the rate of evolution of the period distribution strongly depends on the initial cluster concentration. The larger the concentration, the larger the change in the period distribution. For more strongly concentrated clusters,the period distribution after about 10 Myr of cluster evolution is similar to the period distribution after a few Gyr of evolution in an initially tidally filling cluster, known as the density degeneracy (Marks \& Kroupa 2012, Marks et al. 2014). The maximum of the period distribution moves towards smaller periods as the cluster ages and the distribution shape becomes quickly bell-shaped, as observed for Galactic field populations (e.g. Raghavan 2010 for solar-type stars). The initial change in the distribution is rapid, occurring on a crossing-time scale (Marks, Kroupa \& Oh 2011), after which the cluster enters a phase of slow, two-body relaxation driven binary processing. At 12 Gyr, the maximum of the period distributions for the tidally under-filling and the tidally filling models are about $10^3$ days and $10^6$ days, respectively. 

\begin{figure}[h]
 \vspace*{-0.1 cm}
\begin{center}
 \includegraphics[width=7.0cm, angle=-90]{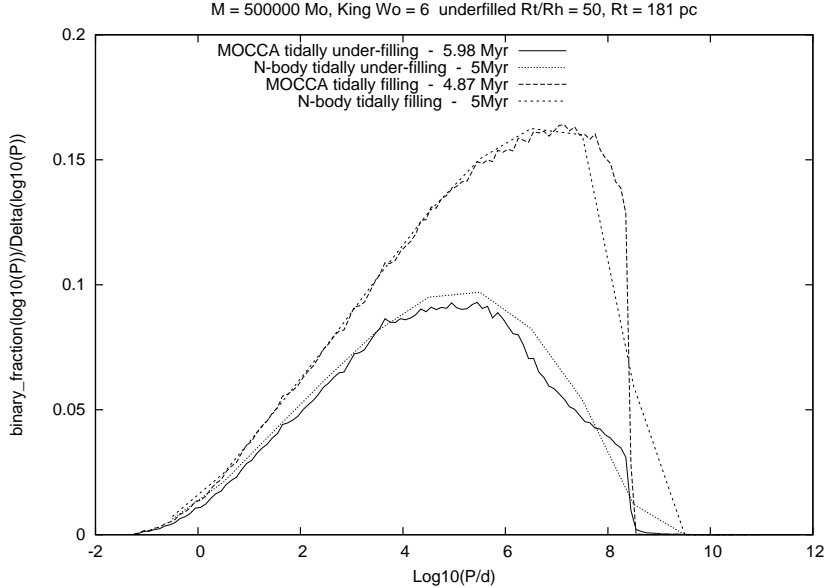} 
 \vspace*{0.8 cm}
 \caption{The binary period distribution for tidally filling and tidally under-filling models for MOCCA and N-body (using the semi-analytical BiPoS code) after 5 Myr computation time.}
   \label{fig2}
\end{center}
\end{figure}

Fig. 2 shows a comparison between the binary distributions at about 5 Myr for the tidally filling and tidally under-filling MOCCA models and the BiPoS code. The MOCCA and N-body period distributions agree remarkably well, both for the tidally 
filling and under-filling models, and so do the distributions for binary binding energy, mass ratio and eccentricity. This is somewhat unexpected, since Marks, Kroupa \& Oh (2011) find that very fast destruction of binaries takes place on a dynamical time scale and strongly depends on the cluster concentration. The larger the cluster concentration, the larger the rate of binary destruction. Processes with characteristic time scales much shorter than the local relaxation time, and comparable to the dynamical time scale, are in principle not well followed by the Monte Carlo method. And yet, the good agreement between BiPoS and MOCCA suggests that the probability of binary dynamical interactions occurring in MOCCA is properly computed (taking into account the local cluster properties), and so the binary destruction rates are well reproduced. In the MOCCA simulations, some cooling of the system is observed (a small decrease in $R_h$) initially because of binary disruption (up to a time of 5 Myr), but then heating of the system takes over due to binary hardening.

\subsubsection{Comparison to Observed Properties in Globular Clusters}

The results of comparing the observational data from Milone et al. (2012) to the results of our MOCCA simulations are discussed in great detail in Leigh et al. (2014). Here we will only summarize the main conclusions. Only the tidally under-filling models can reproduce the observations. For these models, both the binary fractions and the ratio of the binary fractions inside and outside $R_h$  reproduce the observations reasonably well. The tidally filling models, however, failed to reproduce the observations. For these models, a correlation is observed between the cluster mass and the binary fraction outside $R_h$, instead of an anti-correlation. If the initial binary populations are indeed described by the Kroupa (1995) distribution, then our comparison of the MOCCA computations with Milone et al. (2012) suggests that globular clusters must have formed strongly tidally under-filling. This is necessary to create sufficient dynamical processing to reproduce the observed anti-correlation between the binary fraction outside $R_h$ and the total cluster mass. It is worth noting that those MOCCA simulations with high initial concentrations, initial binary fractions of about 10\% and a flat distribution in the logarithm of the semi-major axis cannot recover the observed anti-correlation for the binary faction outside $R_h$ (Leigh et al. 2013b). Of course, this does not rule out other types of distributions of the primordial binary parameters, which might also provide a good fit to the observed binary properties. 

\subsection{A New IMBH Formation Scenario}

As was already mentioned in Section 3, the new scenario for IMBH formation came as a byproduct of projects carried out with Nathan Leigh and collaborators (Leigh et al. 2013b, Leigh et al. 2014). While working with the data produced during the simulations done for the first project, we noticed rather unexpectedly that for some models a slow buildup of BH mass is observed. 

\begin{figure}[h]
 \vspace*{-0.1 cm}
\begin{center}
 \includegraphics[width=7.0cm, angle=-90]{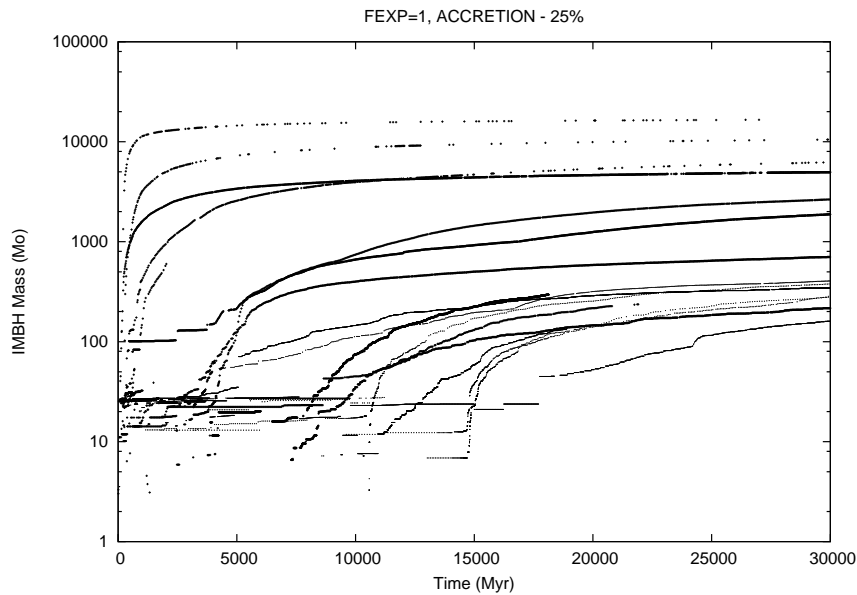} 
 \vspace*{0.8 cm}
 \caption{IMBH mass build up for models with reduced mass accretion onto the IMBH (25\% of the standard BSE setup) and reduced standard star expansion after merger events ($1/3$ of the standard Fewbody setup). See description in the text.}
   \label{fig3}
\end{center}
\end{figure}

As is illustrated in Fig. 3, there are two different regimes of BH mass buildup. The first starts later on in the cluster evolution and has a rather small rate of mass increase (SLOW scenario), while the second starts very early on in the cluster evolution with a very high rate of mass increase (FAST scenario). The BH mass buildup is observed for runs with and without mass fallback (Belczynski et al. 2002) and, what is very unexpected, also for simulations done for massive and dense open clusters (project with Christoph Olczak - work in progress). Generally, only a small fraction of all models show a significant buildup of BH mass, and hence IMBH formation. The process of IMBH formation is highly stochastic. A quick inspection of the runs allows us to formulate the following rules: the larger the initial cluster mass, the larger the probability of IMBH formation; the larger the initial concentration, the larger the probability of IMBH formation; the larger the initial concentration, the earlier and faster an IMBH is formed. 

Most of our simulations were carried out with the assumption that 100\% of the mass of a star colliding with a BH is accreted onto the BH, and the final size of an object formed in a collision is three times larger than the sum of the radii of the colliding stars ($F_{exp}=3$). These are the standard assumptions applied in the BSE code (Hurley et al. 2000, 2002) and in the Fewbody code (Fregeau et al. 2004). As it was pointed out during the MODEST-14 conference, these assumptions might be too strong. The process of mass accretion onto the BH is very complicated, and recent simulations suggest that less than 50\% of the incoming star mass is directly accreted onto the BH. So, in the next set of simulations, we weaken the standard assumptions - only 25\% of the mass of the incoming star is accreted, and the size of the final merged object is just the sum of the radii of the colliding stars ($F_{exp}=1$). As is clear from Fig. 3, an IMBH still forms. As expected, however, the IMBH formation is less efficient. Nevertheless, this strengthens the evidence in favor of our new scenario for IMBH formation. 

\begin{figure}[h]
 \vspace*{-0.1 cm}
\begin{center}
 \includegraphics[width=7.0cm, angle=-90]{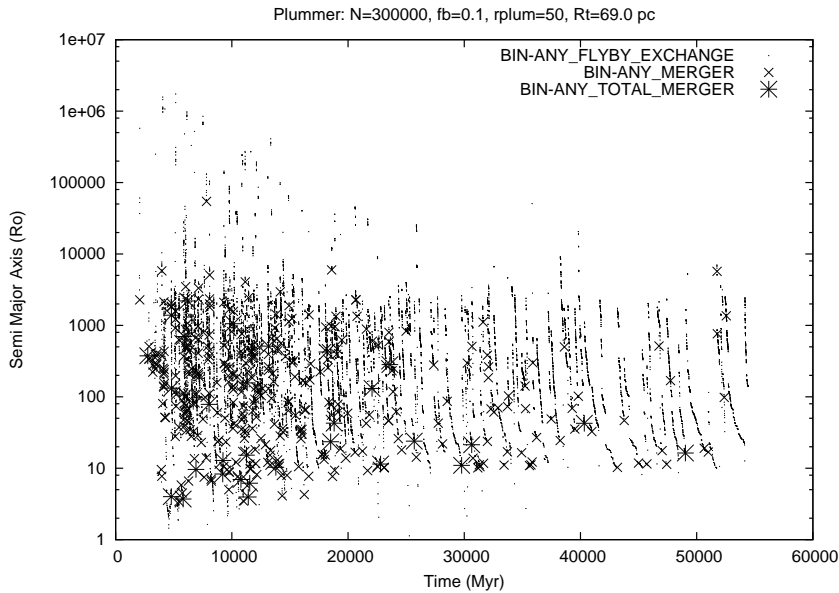} 
 \vspace*{0.8 cm}
 \caption{Evolution of the semi-major axis for binaries containing an IMBH. Symbols: plus - flyby and exchange, cross - merger, star - total merger.}
   \label{fig4}
\end{center}
\end{figure}

Fig 4 shows the evolution of the semi-major axes of binaries containing an IMBH for the SLOW scenario. Clear patterns in the subsequent binary evolution are apparent - e.g. shrinkage of the binary semi-major axis (binary hardening). This is due to dynamical interactions with other binaries and single stars, as well as binary evolution connected either with stellar evolution or gravitational wave radiation (GR). Binary mergers resulting purely from binary evolution are rare. More frequent are binary mergers connected with dynamical interactions. There are two kinds of dynamical mergers: mergers with one binary component (which preserve the binary), and total mergers in which all interacting stars merge into one object. The total binary merger events are crucial from the point of view of IMBH formation. Because of these interactions, an IMBH binary (when its mass is still low) cannot harden to the point of being able to escape from the cluster, which can occur if it receives a strong recoil during a subsequent dynamical interaction. So, the IMBH remains in the system and, consequently, is able to steadily grow in mass. 

In the case of the SLOW scenario, the central cluster densities are not very high - only $10^5 M_{\odot}/pc^3$. Thus, a special set of initial conditions is not needed to form an IMBH. The situation is different for the case of the FAST scenario. The densities required for significant IMBH mass buildup to occur are very high, greater than $10^8 M_{\odot}/pc^3$. These extremely high densities are needed when BHs form a bound and very dense subsystem in the cluster center - mergers of binaries with BHs have to be more efficient than the removal of BHs from the cluster due to strong recoils in dynamical interactions. Such high densities are not very probable in the GCs observed in the Milky Way, but they can occur in nuclear star clusters (NSC). Perhaps the FAST scenario for IMBH formation discussed here occurs commonly in the NSCs of low-mass galaxies.  

The new scenario for IMBH formation can be summarized as follows:
 
\begin{itemize}
\item To initiate the process of BH mass growth, either at least one BH must be left in the cluster after the early phase of SN explosions, or a single BH must be formed via mergers during dynamical interactions. If several BHs remain in the system, the cluster density has to be extremely high for an IMBH to form, greater than  $10^8 M_{\odot}/pc^3$;
\item Next, the formation of a BH-any star binary forms via three-body interactions. The BH is the most massive object in the cluster, so there is a high probability that the BH will be exchanged into, or form a binary. Frequently, BH companions are main sequence (MS), red giant (RG) or asymptotic giant branch (AGB) stars (possibility for X-ray emission); 
\item Dynamical interactions with other binaries and stars:
\begin{itemize}
\item orbit tightening leading to mass transfer from MS/RG/AGB companions;
\item exchanges and mergers, leaving the binary in tact;
\item total mergers in dynamical interactions or the emission of gravitational waves - in this case, the binary is destroyed and only the BH is left. The single BH is then formed a new binary via another three-body interaction, which is free to undergo subsequent dynamical interactions with other single and binary stars.  In this way, the BH mass steadily increases.
\end{itemize}
\end{itemize}

\begin{figure}[h]
 \vspace*{-0.1 cm}
\begin{center}
 \includegraphics[width=7.0cm, angle=-90]{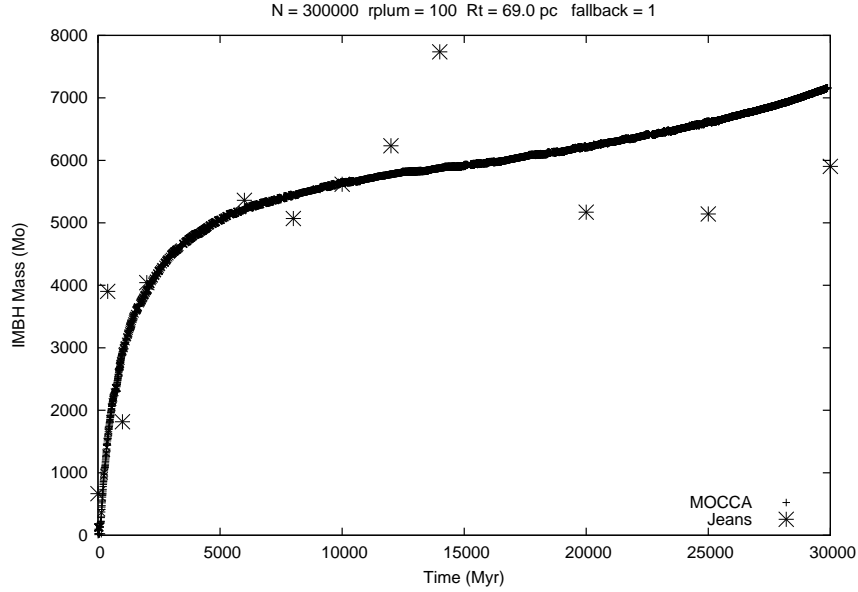} 
 \vspace*{0.8 cm}
 \caption{IMBH mass from our MOCCA simulations and from fitting Jeans models (Lü\"{u}tzgendorf et al. 2013) to the SBP and VDP obtained from the MOCCA simulations.}
   \label{fig5}
\end{center}
\end{figure}

The buildup of the IMBH mass for a model with N=300000, $R_{plum} = 100$ and mass fallback for SN natal kicks is shown in Fig 5. The mass of the IMBH increases up to 7000 $M_{\odot}$. The surface brightness profiles (SBP) and velocity dispersion profiles (VDP) constructed from this simulation were used by Nora Lü\"{u}tzgendorf to fit the IMBH mass with her code based on Jeans' model (Lü\"{u}tzgendorf et al. 2013). As seen in Fig. 5, the fit to the MOCCA data is rather good. MOCCA is able to reproduce more or less correctly the system structure around an IMBH. This is a rather unexpected result, given that MOCCA was not designed to model the relevant physics near a massive IMBH. In fairness though, the agreement is only good in some of our models. Definitely, more work is needed.

\begin{figure}[h]
 \vspace*{-0.1 cm}
\begin{center}
 \includegraphics[width=7.0cm, angle=-90]{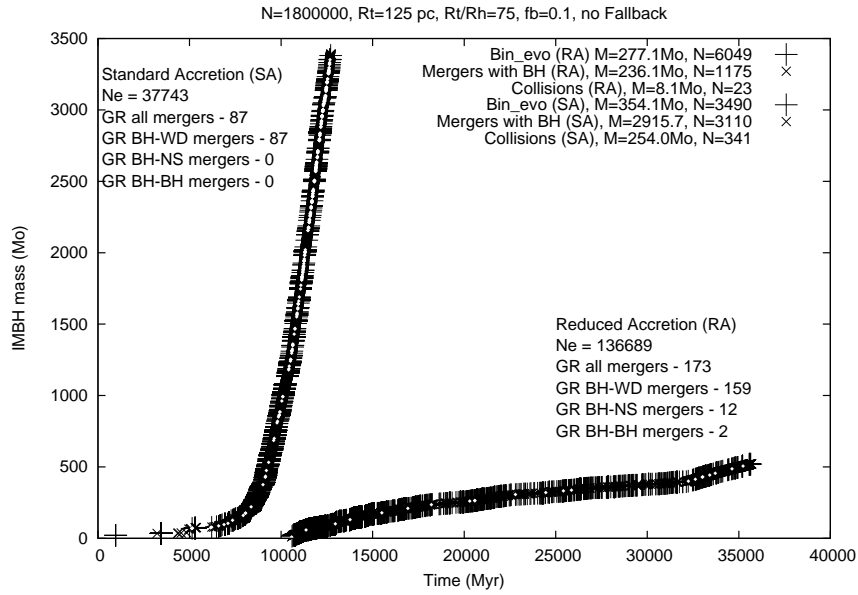} 
 \vspace*{0.8 cm}
 \caption{IMBH mass buildup for models with standard and reduced mass accretion onto the IMBH. Mass transfer events, mergers and collisions with the IMBH are depicted by different symbols (see insets). Also listed are the numbers of interactions, and the total mass gained by the IMBH in each dynamical interaction.}
   \label{fig6}
\end{center}
\end{figure}

Fig. 6 shows the number of events in which mass is transferred onto an IMBH, either from its companion during binary evolution, or because of collisions with incoming stars for the SLOW scenario for standard and reduced accretion. Such mass accretion events are potentially observable because of associated electromagnetic or gravitational radiation. The number of events for collisions and GR are listed in Fig. 6. The numbers are substantial. Interestingly, there are more GR and binary evolution events for the case of reduced mass accretion onto the IMBH. This is connected to the fact that mergers during binary dynamical interactions produce far fewer "total" mergers, due to the smaller size and hence cross-section of the merged object. Instead, there is a large number of GR events, mainly associated with BH-WD mergers. For the FAST scenario, physical collisions are the dominant type of interaction provided the IMBH is sufficiently massive (larger than 5000 - 7000 $M_{\odot}$). 

Let's assume that the probability of IMBH formation depends mainly on the average binary interaction probability. Then we can compare models with and without IMBH formation to see if they occupy separate regions in Escape Velocity - Interaction Probability space. To calculate the interaction probability, we assume an interaction involving an average binary at the soft/hard boundary and an average single star, that occurs inside $R_h$. Fig. 7 shows the results of this exercise. Indeed, at time T=0, there is a statistical boundary above which the probability for IMBH formation is substantial. This boundary line is drawn by eye. Models with substantially reduced mass accretion are only likely to form an IMBH for cluster escape velocities larger than about 30-40 km/s. The same exercise is repeated at time T=12 Gyr (i.e. at the present-day), which shows that indeed some Galactic GCs occupy a region in Escape Velocity -Interaction Probability space for which the MOCCA models do produce an IMBH. These clusters include Omega Cen, 47Tuc, M22 and NGC6293. 

\begin{figure}[h]
 \vspace*{-0.1 cm}
\begin{center}
 \includegraphics[width=7.0cm, angle=-90]{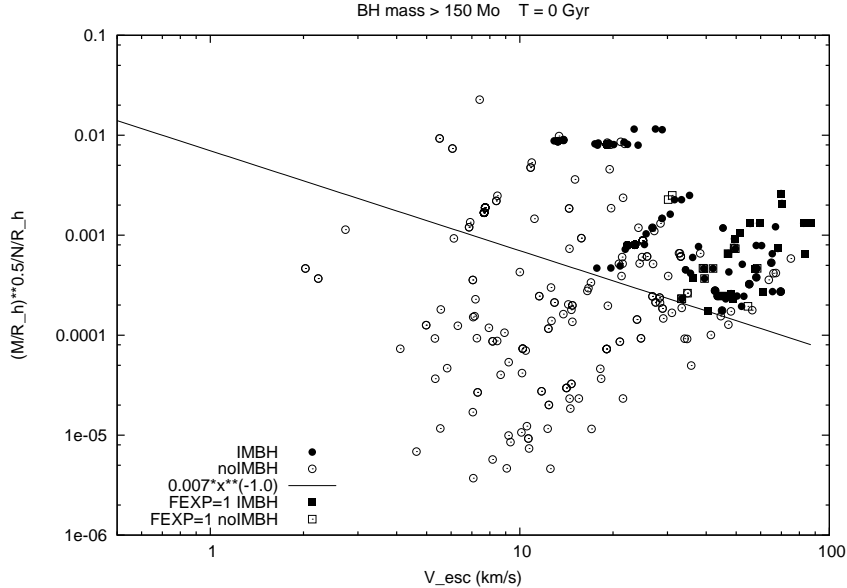} 
 \vspace*{0.8 cm}
 \caption{The interaction probability vs escape velocity at time T=0 for all models (see the inset).}
   \label{fig7}
\end{center}
\end{figure}

Concluding, it is worth mentioning that models with reduced natal kicks (because of mass fallback) for BHs may still contain a substantial number of stellar mass BHs even after a Hubble time of cluster evolution. The number of retained BHs depends on the cluster mass and concentration, via the cluster half-mass relaxation time. The larger the half-mass relaxation time, the larger the number of retained BHs. This means that those clusters most likely to host stellar mass BHs are massive and have a low concentration, instead of massive and dense. Also, simulations of dense and massive GCs show that NSs and BHs can form in substantial numbers not only due to supernovae explosions, but also because of physical collisions and binary mergers later on in the cluster evolution. Interestingly, IMBH formation suppresses the production of any binaries with NS or BH companions, and even the formation of NSs and BHs themselves. This is because their progenitors (white dwarfs, NSs and BHs) are very quickly removed from the system via dynamical interactions with the IMBH. They are the most massive stars in the cluster, and therefore the probability for their involvement in dynamical interactions is high.

\section{Conclusions}

Here are our conclusions, which summarize the talk.
\begin{itemize}
\item The MOCCA code is able to follow the initial destruction of wide binaries. In support of this, the results are in good agreement with the semi-analytic code BiPoS, which is based on N-body simulations.
\item If the initial binary population is described by the Kroupa (1995) 
distributions, then our comparison of the MOCCA computations with Milone et al. (2012) suggests that globular clusters need to have formed strongly 
tidally under-filling to produce sufficient dynamical processing, and hence to reproduce the observed anti-correlations between the binary fractions and the total cluster mass.
\item NSs and BHs can form (in substantial numbers) in the course of star cluster evolution due to dynamical interactions (collisions and binary interactions);
\item If the cluster density is large enough (about $10^5 M_{\odot}/pc^3$), an existing BH can experience substantial mass buildup due to collisions/mergers during dynamical interactions and mass transfer in binaries. The SLOW scenario is more probable;
\item The process of BH mass buildup, and finally IMBH formation, is highly stochastic. The rate of IMBH mass buildup strongly depends on the cluster density. The larger the density, the higher the rate;
\item There are frequent phases of mass transfer in binaries containing an IMBH and mergers of stars with an IMBH. Therefore, X-ray emissions and/or GR could be observable during these events. 
\end{itemize}

Of course one should be aware about possible problems associated with the approximations used in the MOCCA code.
\begin{itemize}
\item The MOCCA code does not cope well with physical processes that have a characteristic time scale comparable to the dynamical time scale; 
\item The MOCCA code is not prepared to follow the dynamical evolution of extremely massive objects (larger than a few hundred $M_{\odot}$). Nevertheless, the initial IMBH mass buildup is modeled correctly;
\item The Fewbody code (Fregeau et al. 2004) seems to work properly for extreme mass ratios - several checks were carried out; 
\item There are some doubts about the accuracy of the BSE code (Hurley et al. 2000, Hurley et al. 2002), and its ability to follow binary evolution and mass transfer involving extreme mass ratios and massive compact objects. The mass transferred onto an IMBH because of binary/stellar evolution is not the dominant source of mass accretion, so the process of IMBH mass buildup can occur even if binary evolution-induced mass transfer is completely switched off;
\item Very large kicks associated with the mergers of BHs having misaligned spin vectors.
\end{itemize}

\vskip 0.5cm
{\bf{ACKNOWLEDGMENTS}}

AH, MG and AA were partly supported by the Polish Ministry of Science and Higher Education and by the National Science Centre through the grants  DEC-2011/01/N/ST9/06000 and DEC-2012/07/B/ST9/04412, respectively.  NL is thankful for the generous support of an NSERC Postdoctoral Fellowship.

\end{document}